\documentclass{article}
\renewcommand{\P}[2]{{\cal E}\mbox{$[#1,#2]$}}

\begin{document}

\title{
A Probabilistic Analysis of the Power of Arithmetic Filters\thanks{
  \em 
  This work was partially supported by
  ESPRIT LTR 21957 (CGAL) and by
  the U.S. Army Research Office under grant DAAH04-96-1-0013.
  This work was done in part while O. Devillers was visiting Brown University.
} }

\author{
Olivier Devillers\thanks{INRIA, BP 93, 06902 Sophia Antipolis, France.
Olivier.Devillers@sophia.inria.fr}
\and Franco P. Preparata\thanks{F. Preparata address is Brown Univ.,
Dep. of Computer Science, Providence, RI 02912-1910 (USA). franco@cs.brown.edu}
}
\maketitle

\abstract{      
The assumption of real-number arithmetic, which is at the basis of conventional
geometric algorithms, has been seriously challenged in recent years,
since digital computers do not exhibit such capability.
 A geometric predicate  usually consists of evaluating
the sign of some algebraic expression.
In most cases, rounded computations yield a reliable result, but
sometimes rounded arithmetic introduces errors which may invalidate
the algorithms. The rounded arithmetic may produce an incorrect result
only if the exact  absolute value of the algebraic expression is smaller
than some (small) $\varepsilon$, which  represents the largest error
that may arise in the evaluation of the expression. The threshold 
$\varepsilon$ depends on the  structure of the expression and on the
adopted computer arithmetic, assuming that the input operands are error-free.
 A pair (arithmetic engine,threshold)  is an {\em arithmetic filter}.
In this paper we develop a general technique for assessing the efficacy
of an arithmetic filter. The analysis consists of evaluating both the
threshold and the probability of failure of the filter.
 To exemplify the approach, under the assumption that the input points be 
chosen randomly in a unit ball or unit cube with uniform density, 
we analyze the two important predicates "which-side'' and 'insphere''.
We show that the probability that the absolute values of the corresponding
determinants be no larger than some positive value $V$,
with emphasis on small $V$, is $\Theta(V)$ for the which-side predicate,
while for the insphere predicate it is $\Theta(V^{\frac{2}{3}})$ in dimension 1,
$O(V^{\frac{1}{2}})$ in dimension 2,
and $O(V^{\frac{1}{2}}\ln \frac{1}{V})$ in higher dimensions.
Constants are small, and are given in the paper.
}

\section{Introduction}
 
 The original model of Computational Geometry rests on real-number
arithmetic, and under this assumption the issue of precision is
irrelevant. However, the reality that computer calculations have
finite precision has raised an increasing awareness of its effect on the
quality and even the validity of geometric algorithms conceived within
the original model, in the sense that algorithm correctness does not
automatically translate into program correctness. In recent years
this issue has been amply debated in the literature (see, e.g.,
\cite{bkmnsu-egcl-95,fv-eeacg-93,y-tegc-97}).
In particular, it has been observed that while some degree of
approximation may be tolerated in geometric constructions, the
evaluation of predicates ( the "tests " carried out in the execution
of programs, -- such as which-side, incircle, insphere --)
must be exact to ensure the structural (topological)
correctness of the results \cite{bms-hcvdl-94,y-tegc-97,lpt-rpqiv-96}.
 
In principle, error-free predicate evaluation
is achievable for error-free input operands, if the latter are
treated as integers and the arithmetic is carried out with whatever
operand length is required to express the intermediate results.
Such safe approach, however, if adopted in
its crudest form, would involve enormous overheads
and would be nearly impracticable,
since the execution time of some operation (such as multiplication)
may increase quadratically
with the length of the representation.
 
 As a time-saving alternative to exact arithmetic, it has been
customary to resort to rounded (approximate) arithmetic (for example,
floating-point arithmetic). Such practice can be modeled as follows.
Evaluation of  a predicate $P$ typically involves computing the value
$\mu$ of some expression, built using rational operations.
The value of $\mu$ can be mapped to  one of three values: positive, zero,
negative ( referred to  here as the "sign"
of $\mu$), which defines the predicate $P$.
Let ${\cal E}$ denote an evaluator for  $P$, and let $\mu ({\cal E})$ denote
the numerical value computed by ${\cal E}$. In general,
${\cal E}$ is an approximate evaluator of $\mu$, so that its use
involves the adoption of a device, called a {\em certifier}, intended
to validate the correctness of the evaluation. The pair
(evaluator, certifier) is what has been refered to as a {\em filter} in
the literature \cite{fv-eeacg-93,mn-iga-94}. 
Typically, the certifier for ${\cal E}$ compares $|\mu ({\cal E})|$ with
a fixed threshold $\varepsilon({\cal E}) \geq 0$.
If $|\mu ({\cal E})| \geq \varepsilon ({\cal E})$,
then the sign of $\mu ({\cal E})$ is reliable.
Otherwise, the certifier is unable to validate the result and
we have a {\em failure} of the filter. In such event
recourse to a more powerful filter is in order.
This suggests the need to develop a family of filters of increasing
precision (and complexity), to be used in sequence until
failure no longer occurs. The last item of this sequence is
the exact evaluator, for which the certifier is vacuous (i.e.,
$\varepsilon({\cal E}) =0$). Such approach,  with an
obvious trade-off between efficacy and efficiency, embodies the
notion of {\em adaptive precision}.
 
From a practical standpoint it is therefore very important to
gauge the efficacy of very simple filters, that is,
their probability of success. If it turns out that if a filter
has a high probability of success, then recourse to  a more time-consuming
filter (or exact computations) will be a rather rare event.
Of course, any such estimate of efficacy rests on some arbitrary
hypotheses on the {\em a priori} probability of problem instances.
This is an important  {\em caveat}; however, under reasonable hypotheses
(uniform distributions), we submit that the obtained estimates
will be a significant contribution to the assessment of the validity of
such approaches.
 
In this paper we analyze two specific predicates,
``which-side'' and ``insphere''.
Both predicates consist of computing the signs of appropriate
$\delta \times \delta$ determinants, whose entries are specified with
a fixed number of bits. Depending upon the adopted evaluation scheme
(the choice of an equivalent expression for a given function) and
upon the precision of the input operands (for example, only a fixed-length
prefix of their representation may be used in the evaluation),
only a prefix of the computed value is reliable.
This means that if the absolute value of the
determinant is above a known threshold $\varepsilon$, then its sign is also
reliable.

Our objective is therefore two-fold:
\begin{enumerate}
\item To compute the value of the threshold $\varepsilon$
for a given determinantal evaluation technique;
\item To compute the probability that the absolute value of
the result of the evaluation does not exceed $\varepsilon$, i.e.,
the probability of filter failure.
\end{enumerate}
 
We recall that when evaluating the determinant
of a $\delta \times \delta$ matrix $A$
(a $\delta$-determinant, for short) we are computing the {\it signed} measure of a
hyperparallelepiped defined by the $\delta$ vectors corresponding to the rows of
$A$. Since each of the components of these vectors is an integer in
the range $[-2^{b-1}, 2^{b-1}-1]$, a generic vector is (applied to the
origin and) defined by its free terminus
at grid points in a $\delta$-dimensional cube
${\cal C}_{\delta}$
of sidelength $2^b$ centered at the origin.
 
Our probability model assumes that all
grid points within ${\cal C}_{\delta}$ have identical
probability. Our analysis aims at estimating (a majorization of)
the distribution of the volume of the hyperparallelogram described above.
To obtain the desired result, we introduce some simplifications
consistent with the objective to majorize the probability.
Specifically, while our assumption is a discrete uniform
distribution within ${\cal C}_{\delta}$, we begin by considering
a continuous uniform distribution within the ball ${\cal B}_{\delta}$,
the $\delta$-dimensional ball of radius 1.
The obtained results are then used to bound from above the distribution
of the volume for uniform density within the cube
${\cal C}_{\delta}$ (the $\delta$-fold cartesian product
of interval $[-1,1]$), and are finally extended to the target case
of uniform discrete distribution.
We shall recognize that the initial simplification (uniform density in
${\cal C}_{\delta}$) closely approximates the more realistic situation.
 
The paper is organized as follows.  We begin with the "which-side''
predicate (referred to as "determinant''), i.e., in Section 
\ref{sec_determinant} we carry out the
probabilistic analysis in ${\cal B}_{\delta}$, for $\delta= 1,2,
3$ and arbitrary $\delta$ (detailed considerations of the low-dimensional cases
has an obvious pedagogical motivation). In Section \ref{sec_cube}, we extend the
results from the continuous ball to the discrete cube. In Section
\ref{sec_sphere}
we carry out an analogous analysis for the "insphere'' predicate,
which illustrates the adverse  effect of dependencies among
the determinant entries. Finally, in Section \ref{application}
we evaluate the
precision of determinant evaluation by recursive expansion, and
illustrate the efficacy/efficiency tradeoff.

\section{Probabilistic analysis within unit ball\label{sec_determinant}}

Throughout this section we  adopt the following notation: 
We let $x_1,x_2,\ldots, x_\delta$ be the coordinates of $\delta$-space,and 
$p_1,p_2,\ldots p_{\delta}$  be $\delta$  points in the
unit ball of dimension $\delta$. We also denote by
$|p_1,p_2\ldots p_{\delta}|$ the absolute value of the determinant defined
by points $p_1,p_2\ldots p_{\delta}$.   This quantity, which is
the volume of the hyperparallelogram
defined by the origin and by points $p_1,p_2\ldots p_{\delta}$, will also be
denoted $a_{\delta}$.

We begin by examining in some detail the
cases of low-dimension determinants. Since the analysis is done in a
visualizable geometric setting ($\delta \leq 3$), it is preparatory to 
the more abstract higher-dimensional cases.

\subsection{ 1-and 2-determinant}
Obviously, if $p_1$ is uniformly distributed between -1 and 1, then
 
\[ Prob(|p_1|\leq R)=R \]

Less trivial is the analysis of the two-dimensional case.
We will study the probability for $|p_1,p_2|$ to be smaller than a constant $A$
when $p_1$ and $p_2$ are distributed uniformly in the unit disk.

\begin{figure}[t]
\begin{center}
\def\IPEfile{Det2.ipe}\input{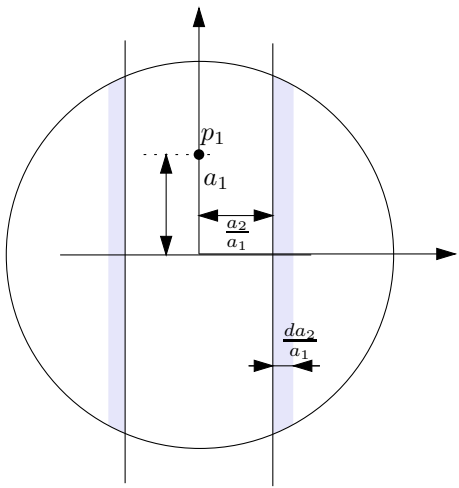}
\caption{For the analysis of the 2-determinant\label{Det2}}
\end{center}
\end{figure}

Once $p_1$ is chosen (due to the circular symmetry, $p_1$ is represented
by a single parameter $a_1$, its distance from the origin),
$p_2$ will yield an area between $a_2$ and $a_2+da_2$ if
it belongs to one of the  two strips  of width $\frac{da_2}{a_1}$ depicted
in Figure  \ref{Det2}.

We then have

\begin{eqnarray*}
Prob(|p_1,p_2|\leq A)
&=& \int_0^1 Prob( |p_1,p_2| \leq A \mid a_1) p(a_1) da_1.
\end{eqnarray*}

\noindent Since the density function of $a_1$ is  
$p(a_1)=\frac{2\pi a_1}{\pi}=2a_1$,
and the density of $a_2$ conditional on  $a_1$  is 
  $p(a_2\mid a_1)da_2=\frac{1}{\pi}4\sqrt{1-\left(\frac{a_2}{a_1} \right)^2}\frac{da_2}{a_1}$ 
($(\frac{1}{\pi})$ is the density in the unit disk, and
$4\sqrt{1-\left(\frac{a_2}{a_1} \right)^2}$
is the total length of the two strips)
we have

\begin{eqnarray*}
Prob(|p_1,p_2|\leq A)
&=& \int_{a_1=0}^1 \int_{a_2=0}^{\min(A,a_1)} 2p(a_2|a_1)a_1 da_1 da_2\\
&=& \int_0^A \left( \int_{a_2}^1  \frac{4}{\pi a_1} \sqrt{1-\left(\frac{a_2}{a_1}\right)^2}
						   \cdot 2a_1da_1 \right) da_2\\
&=& \int_0^A \frac{8}{\pi}da_2
        \left| \sqrt{{a_1}^2-{a_2}^2} -a_2 \arccos \frac{a_2}{a_1} \right|_{a_2}^1\\
&=& \int_0^A \frac{8}{\pi}da_2
        \left( \sqrt{1-a_{2}^2} -a_2 \arccos a_2 \right)\\
&=& \frac{8}{\pi} \left| \frac{3}{4}a_2\sqrt{1-a_{2}^2} + \frac{1}{4}
		\arcsin a_2 -\frac{1}{2}a_2^2\arccos a_2 \right|_0^A\\
&=& \frac{6}{\pi} A\sqrt{1-A^2}
   +\frac{2}{\pi} \arcsin A
   -\frac{4}{\pi}A^2\arccos A\\
\end{eqnarray*}

One can easily verify that the value of
the last expression is 1 for $A=1$, which is the maximum
attainable value for the area of the parallelogram.

\subsection{3-determinant}

Again we assume that points are
uniformly distributed in the unit ball, and compute the
probability that the volume of the parallelepiped
defined by $O, p_1, p_2,$ and $p_3$
is smaller than a constant $V \geq 0$.

To compute this volume, we begin
by considering the parallelogram defined by $O, p_1,$ and $p_2$, 
evaluate its area, and then consider the distance of $p_3$ from
the plane containing the parallelogram.

Distance $Op_1$ is between $a_1$ and $a_1+da_1$ if $p_1$ belongs to a spherical
crown of thickness $da_1$ and area $4\pi a_{1}^2$. Therefore the probability
density of $a_1$ is $p(a_1)= \frac{3}{4\pi} 4 \pi a_{1}^2 =3a_{1}^2$ (note that 
$\frac{3}{4\pi}$ is the density in the unit sphere).

\begin{figure}[t]
\begin{center}
\def\IPEfile{Det3.ipe}\input{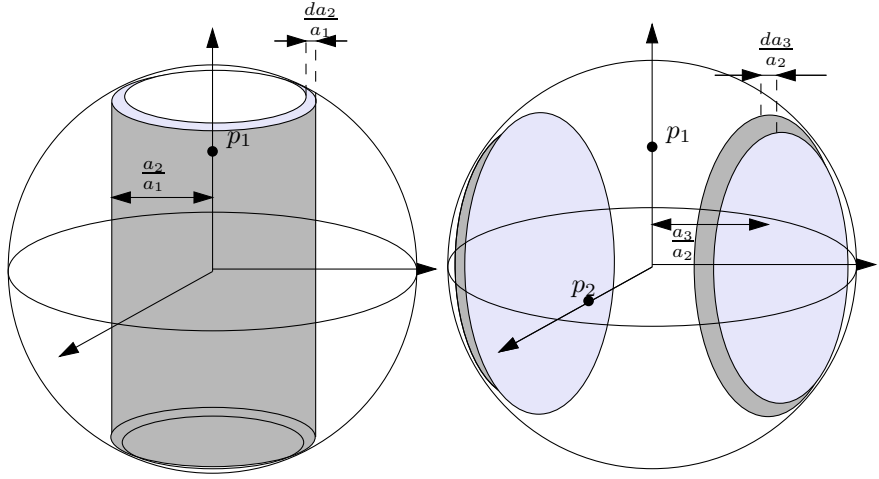}
\caption{For the analysis of the 3-determinant\label{Det3}}
\end{center}	
\end{figure}

Once $p_1$ has been chosen, the area of the parallelogram defined by $O, p_1$
and $p_2$ is between $a_2$ and $a_2+da_2$ if $p_2$ belongs to the crown of
thickness $\frac{da_2}{a_1}$
of a cylinder of radius $\frac{a_2}{a_1}$ whose axis contains
$Op_1$ (see Figure \ref{Det3}a).
Therefore the distribution of $a_2$ conditional on $a_1$ is given
by
\begin{equation}
\label{A2}
p(a_2|a_1)da_2=\frac{3}{4\pi}\cdot 2\pi \frac{a_2}{a_1} \cdot
2\sqrt{1-\left(\frac{a_2}{a_1}\right)^2}\cdot \frac{da_2}{a_1}=
\frac{3a_2}{a_{1}^2}\sqrt{1-\left(\frac{a_2}{a_1}\right)^2}da_2
\end{equation}

Finally, once $p_1$ and $p_2$ have been chosen, the volume of the parallelepiped is
between $a_3$ and $a_3+da_3$ if $p_3$ belongs to one of the two spherical
slices of width $\frac{da_3}{a_2}$,
parallel to the plane containing $O, p_1,$ and
$p_2$, and at distance $\frac{a_3}{a_2}$ from it (see Figure \ref{Det3}b).
Therefore the distribution of $a_3$
conditional on $a_2$ is given by

\begin{equation}
\label{A3}
p(a_3\mid a_2)da_3= \frac{3}{4\pi}\cdot 2\cdot \pi\left(1-\left(\frac{a_3}{a_2}\right)^2\right)\cdot\frac{da_3}{a_2}
= \frac{3}{2a_2}\left(1-\left(\frac{a_3}{a_2}\right)^2\right)da_3
\end{equation}

On the basis of this analysis, we can say
\begin{footnotesize}
\begin{eqnarray*}
\lefteqn{Prob(|p_1,p_2,p_3|\leq V)
}\\
&=&{
 \int_{a_1=0}^1 \int_{a_2=0}^{a_1} \int _{a_3=0}^{\min(V,a_2)}
        p(a_1) p(a_2\mid a_1)p(a_3\mid a_2) da_1da_2da_3
}\\
&=&{
 \int_0^V \left( \int_{a_3}^1
 \left( \int_{a_3}^{a_1} p(a_3\mid a_2)p(a_2\mid a_1) p(a_1)da_2 \right) da_1 \right) da_3
}\\
&=&{
 \int_0^V \frac{27}{2}da_3  \int_{a_3}^1 da_1
	   \int_{a_3}^{a_1}\left( \frac{\sqrt{{a_1}^2-{a_2}^2}}{a_1}
				-{a_3}^2\frac{\sqrt{{a_1}^2-{a_2}^2}}{a_1{a_2}^2}\right)da_2
}\\
&=&{
	 \int_0^V \frac{27}{2}da_3  \int_{a_3}^1 da_1
	 \left(  \left| \frac{a_2}{2} \frac{\sqrt{{a_1}^2-{a_2}^2}}{a_1}
		+ \frac{a_1}{2} \arcsin \frac{a_2}{a_1}  \right|_{a_3}^{a_1}
	  +{a_3}^2\left| \frac{\sqrt{{a_1}^2-{a_2}^2}}{a_2a_1}
			+ \frac{1}{a_1} \arcsin\frac{a_2}{a_1} \right|_{a_3}^{a_1}\right)}\\
&=&{
	 \int_0^V \frac{27}{4}da_3  \int_{a_3}^1 da_1
       \left(   \frac{\pi}{2}a_1
			  + \pi\frac{{a_3}^2}{a_1}
			  -\sqrt{{a_1}^2-{a_3}^2}\frac{3a_3}{a_1}
			  - a_1 \arcsin \frac{a_3}{a_1}
			  - 2\frac{{a_3}^2}{a_1} \arcsin\frac{a_3}{a_1}\right)          }\\
&=&{
	 \int_0^V \frac{27}{4}da_3 \left(
		\frac{\pi}{4}
		-\pi {a_3}^2 \ln a_3
		- \frac{7}{2}a_3\sqrt{1-{a_3}^2}
		+3{a_3}^2\arccos a_3
		-\frac{1}{2}\arcsin a_3
		   -2{a_3}^2 \int_{a_3}^1 \frac{1}{a_1} \arcsin\frac{a_3}{a_1} da_1 \right) }\\
&=&{
	\left|         
                        \frac{27\pi}{16}a_3
			-{\frac {9\pi}{4}} {a_3}^{3}\ln (a_3)
				+{\frac {3\pi}{4}} {a_3}^{3}
			+{\frac {63}{8}} \sqrt{1-{a_3}^{2}}^3
			+{\frac {27}{4}} {a_3}^{3}\arccos(a_3)
   				-{\frac {27}{4}} {a_3}^{2}\sqrt {1-{a_3}^{2}}
				-{\frac {9}{2}} \sqrt{1-{a_3}^{2}}^3
\right.}\\ \lefteqn{\hspace*{2cm} \left.
			-{\frac {27}{8}} a_3\arcsin(a_3)
				-{\frac {27}{8}} \sqrt {1-{a_3}^{2}}
         \right|_0^V
            - \frac{27}{2}  \int_0^V \int_{a_3}^1 \frac{{a_3}^2}{a_1} \arcsin\frac{a_3}{a_1} da_1 da_3} \\
&=&{
	 \left|
                        \frac{27\pi}{16}a_3
                         - \frac{9\pi}{4} {a_3}^3 \ln a_3
                         + \frac{3\pi}{4} {a_3}^3
                         + \frac{27}{4} {a_3}^3 \arccos a_3
                         -\frac{81}{8} {a_3}^2 \sqrt{1-{a_3}^2}
                         -\frac{27}{8}a_3 \arcsin a_3
                      \right|_0^V
}\\ \lefteqn{\hspace*{2cm}
            - \frac{27}{2}  \int_0^V \int_{a_3}^1 \frac{{a_3}^2}{a_1} \arcsin\frac{a_3}{a_1} da_1 da_3} \\
\end{eqnarray*}
\end{footnotesize}

The latter integral is not elementarily computable. Since the integrand is
always positive, so is the integral.
To neglect it corresponds to majorizing the probability, which is
conservative for our analysis.
Therefore we write
\begin{scriptsize}
\[
Prob(|p_1,p_2,p_3| \! \leq \! V)
\leq
                \frac{27\pi}{16} V
                - \frac{9\pi}{4} V^3 \ln V
                +\frac{3\pi}{4}V^3
                -\frac{81}{8}V^2\sqrt{1-V^2}
                + \frac{27}{4} V^3\arccos V
                - \frac{27}{8} V \arcsin V\\
\]
\end{scriptsize}

If we set $V=1$, then  the neglected term can be evaluated exactly to
$\frac{3\pi}{4}-1$
by exchanging the order of integration,
and we correctly obtain the value 1 for the probability.

Furthermore, using the inequalities $a_3 \leq a_1$
and $\arcsin\frac{a_3}{a_1} \leq \frac{\pi}{2}$,
the neglected term can be bounded from above as
$\frac{27}{2}  \int_0^V \int_{a_3}^1 {a_3} \frac{\pi}{2} da_1 da_3
\leq \frac{27\pi}{8} V^2$, so we  guarantee the tightness
of the approximation of the probability by $ \frac{27\pi}{16} V$
when $V$ is small.

The preceding analysis, in its simplicity, reveals the essential items
for the evaluation of the relevant conditional probability densities.
Specifically, referring concretely to the case $\delta=3$,
due to the assumption of uniform
distribution of the points in the unit sphere, the conditional probability density
$p(a_i|a_{i-1})da_i$ of $a_i$ given $a_{i-1}$, $i=2,3$, is proportional
(through the value of the density in the unit sphere) to the 
volume of some three-dimensional domain. The latter is a thin crown
(of thickness $\frac{da_i}{a_{i-1}}$)
of a three-dimensional surface ${\cal S}_{3,i}$
which is the locus of the points
at a distance between $\frac{a_i}{a_{i-1}}$
and $\frac{a_i}{a_{i-1}} + \frac{da_i}{a_{i-1}}$ from
the flat ${\cal F}_{i-1}$ spanned by variables $x_1,\ldots, x_{i-1}$.
Surface ${\cal S}_{3,i}$ has a very simple structure.
Let $(u,v)$ be a pair of points realizing the distance
$\frac{a_i}{a_{i-1}}$,
with $u \in {\cal F}_{i-1}$ and $v \in {\cal S}_{3,i}$. 
Point $v$ belongs to the boundary of a $(4-i)$-dimensional ball of 
radius $\frac{a_i}{a_{i-1}}$
of which point $u$ is the center: therefore this entire
boundary  belongs to ${\cal S}_{3,i}$ ( in the discussion
above, this boundary consists of a circle for $i=2$ and of two
points for $i=3$ ). Moreover, since $u$ belongs to
a flat  ${\cal F}_{i-1}$, any translate	 of $v$ within the unit sphere in  a 
flat parallel to  ${\cal F}_{i-1}$ also belongs to ${\cal S}_{3,i}$.
These translates form the intersection of the unit sphere with a flat at
distance  $\frac{a_i}{a_{i-1}}$ from the center of the sphere, and therefore
are an $(i-1)$-dimensional ball of radius
$\sqrt{1- \left(\frac{a_1}{a_{i-1}}\right)^2}$
(in the discussion above this ball consists of a segment for $i=2$ 
and of a disk for $i=3$).
We conclude that ${\cal S}_{3,i}$ is the cartesian product
of the boundary of a $(4-i)$-dimensional ball of radius
$\frac{a_i}{a_{i-1}}$
( the "boundary'' term) and
of an $(i-1)$-dimensional ball of radius
$\sqrt{1- \left(\frac{a_1}{a_{i-1}}\right)^2}$ 
(the "domain'' term).
The expression for the conditional probability density consists of
four factors: the density within the unit sphere, the measure of the
boundary term, the measure of the domain term, and the thickness of the
crown. These four items (in the order given) are evidenced in ({\ref{A2}})
and ({\ref{A3}}).

\subsection{Higher-dimensional determinant}

 We now extend the preceding analysis to arbitrary dimension $\delta$.
If we assume for a one-dimensional volume (a distance) the conventional degree of 1,
then the volume and the surface of  a $j$-dimensional domain have
respective degrees $j$ and $j-1$. Let $v_j(r)$ and $s_j(r)$ respectively
denote volume and surface of a $j$-dimensional ball of radius $r$.
We recall that\cite[9.12.4.6]{b-g-87}
\[
v_i(r)=\frac{\pi^{\frac{i}{2}}}{\frac{i}{2}!} r^i \mbox{ for $i$ even}\;\;\;
v_i(r)=\frac{2^i\pi^{\frac{i-1}{2}}(\frac{i-1}{2})!}{i!}r^i \mbox{ for $i$ odd}.
\]

The probability density $p(a_1)da_1=Prob (a_1\leq|Op_1|\leq a_1+da_1)$
is obviously given
by $\frac{s_{\delta}(a_1)}{v_{\delta}(1)}da_1$.
Referring next to the observations at the end of the preceding subsection, the
conditional probability
density $p(a_i \mid a_{i-1})da_i$  of $a_i$ after
$p_1,p_2,\ldots, p_{i-1}$ have been chosen (conventionally in the flat
described by coordinates $x_1,x_2,\ldots,x_{i-1}$) to realize the
value $a_{i-1}$, has the following expression:
\[
p(a_i|a_{i-1})da_i=
\frac{1}{v_{\delta}(1)} \cdot s_{\delta-i+1}\left(\frac{a_i}{a_{i-1}}\right) \cdot
v_{i-1}\left(\sqrt{1-\left(\frac{a_i}{a_{i-1}}\right)^2}\right)\cdot
\frac{da_i}{a_{i-1}}
\]

Therefore, since $v_i(r)=v_i(1)\cdot r^i$ and
$s_i(r)=i \cdot v_i(1) \cdot r^{i-1}$, we have

\begin{footnotesize}
\begin{eqnarray*}
\lefteqn{p(a_1) \prod_{i=2}^{\delta} p(a_i|a_{i-1})da_i }\\
& = &
\frac{ \delta v_{\delta}(1) a_1^{\delta-1}}{v_{\delta}(1)}
\prod_{i=2}^{\delta}
 \frac{1}{v_{\delta}(1)}\cdot
 (\delta-i+1) v_{\delta-i+1}(1) \cdot \left(\frac{a_i}{a_{i-1}}\right)^{\delta-i} \! \cdot
 v_{i-1}(1) \cdot \left(\sqrt{1-\left(\frac{a_i}{a_{i-1}}\right)^2}\right)^{i-1}
\! \cdot \frac{da_i}{a_{i-1}}\\
& = &
\frac{\delta ! a_1^{\delta-1}}{v_{\delta}(1)^{\delta-1}}
\left(
\prod_{i=2}^{\delta}  v_{\delta-i+1}(1) v_{i-1}(1)
\right)
\left(
\prod_{i=2}^{\delta}
   \frac{{a_i}^{\delta-i}}{{a_{i-1}}^{\delta -i+1}}\left(\sqrt{1-\left(\frac{a_i}{a_{i-1}}\right)^2}\right)^{i-1} da_i
\right)\\
\lefteqn{\mbox{\normalsize In the rightmost term above the product of the powers of the
$a_i$'s simplifies to $\frac{1}{{a_1}^{\delta-1}}$, so that
                        }} \\
& = &
\frac{\delta ! \left(\prod_{i=1}^{\delta-1}  v_i(1) \right)^2}
	 {v_{\delta}(1)^{\delta-1}}
\prod_{i=2}^{\delta}
   \left(\sqrt{1-\left(\frac{a_i}{a_{i-1}}\right)^2}\right)^{i-1} da_i
\end{eqnarray*}
\end{footnotesize}

The last expression contains
a constant depending only on $\delta$,
which will be denoted
\[
k_{\delta} 
\stackrel{\scriptscriptstyle \Delta}{=}
 \frac{\delta ! \left(\prod_{i=1}^{\delta-1}  v_i(1) \right)^2}
	 {v_{\delta}(1)^{\delta-1}}
\]

Now we can write the probability for the absolute value of
the determinant to be no larger than $V$

\begin{footnotesize}

\begin{equation}
\label{Adelta}
Prob(|p_1,p_2\ldots p_{\delta}| \leq V)
= 
k_{\delta} \int_{a_1=0}^1\!\!\int_{a_2=0}^{a_1}\!\!\!\ldots
			   \int_{a_{\delta-1}=0}^{a_{\delta-2}}\!\!
			   \int_{a_{\delta}=0}^{\min(V,a_{\delta-1})}
   \prod_{i=2}^{\delta}
	 \left(\sqrt{1-\left(\frac{a_i}{a_{i-1}}\right)^2}\right)^{i-1}\!\!\! da_i\\
\end{equation}
\[{
 = k_{\delta} \int_{a_1=0}^1\!\!\int_{a_2=0}^{a_1}\!\!\!\ldots
			   \int_{a_{\delta-1}=0}^{a_{\delta-2}}\!\!
	\left(		   \int_{a_{\delta}=0}^{\min(V,a_{\delta-1})}
	  \left(
			\sqrt{1-\left(\frac{a_{\delta}}{a_{\delta-1}}\right)^2}
      \right)^{i-1} \!\!\! da_{\delta}
    \right)
   \prod_{i=2}^{\delta-1}
	  \left(\sqrt{1-\left(\frac{a_i}{a_{i-1}}\right)^2}\right)^{i-1}\!\!\! da_i}
\]
\end{footnotesize}
We now observe that the integral
\[
   \int_{a_{\delta}=0}^{\min(V,a_{\delta-1})}
      \left(\sqrt{1-\left(\frac{a_{\delta}}{a_{\delta-1}}\right)^2}
      \right)^{i-1} da_{\delta}
\]
is trivially bounded by $V$. Therefore we obtain the following upper bound:
\[
Prob(|p_1,p_2\ldots p_{\delta}| \leq V)
\leq
k_{\delta} V \int_0^1\!\!\ldots\int_0^{a_{\delta-2}}
	\prod_{i=2}^{\delta-1}
		 \left(\sqrt{1-\left(\frac{a_i}{a_{i-1}}\right)^2}\right)^{i-1} da_i
\]
Since the latter integral does not depends on $V$, its value is another
constant, which we denote 
${\cal I}_{\delta}$  
and which depends only on $\delta$.
However, when $V=1$, Equation 
(\ref{Adelta}) 
yields
$Prob(|p_1,p_2\ldots p_{\delta}|\leq 1)=k_{\delta}{\cal I}_{\delta+1}$
and, since $Prob(|p_1,p_2\ldots p_{\delta}|\leq 1)=1$,
we get ${\cal I}_{\delta}=\frac{1}{k_{\delta-1}}$.\\
Finally, we conclude
\begin{eqnarray*}
Prob(|p_1,p_2\ldots p_{\delta}|\leq V)
& \leq &
k_{\delta}{\cal I}_{\delta} V\\
& \leq &
\delta \frac{ v_{\delta-1}(1)^{\delta} }{ v_{\delta}(1)^{\delta -1}} V
\stackrel{\scriptscriptstyle \Delta}{=}
\sigma_{\delta} V
\end{eqnarray*}
For small $\delta$, the bounds are given below.
For $\delta=1,2,3$ the bounds coincide with the values previously found,
that is,
\begin{eqnarray*}
\sigma_1 & =  &1\\
\sigma_2 & =  &\frac{2^3}{\pi}              \approx 2.5 \\
\sigma_3 & =  &\frac{3^3\pi}{2^4}           \approx 5.3 \\
\sigma_4 & =  &\frac{2^{13}}{3^4\pi^2}      \approx 10 \\
\sigma_5 & =  &\frac{3^4 5^5\pi^2}{2^{17}}  \approx 19 \\
\sigma_6 & =  &\frac{2^{24}}{5^6\pi^3}      \approx 35 
\end{eqnarray*}

\section{From continuous ball to discrete cube\label{sec_cube}}
\subsection{From continuous ball to continuous cube\label{sec_cont_cube}} 
The above calculations have been carried out
for points uniformly distributed
inside the $\delta$-dimensional ball of radius 1, referred to as  ${\cal
B}_{\delta}$.
This  assumption may not seem to model the real situation for two reasons:
(i) points manipulated by computers have discrete rather than
continuous coordinates,
and (ii) points are more reasonably assumed to be
uniformly distributed in a cube than in a ball (as in the case when each
coordinate is independently and uniformly selected).
 
We will first show that the previous result relative to
the ball ${\cal B}_{\delta}$ induces a similar
result for uniform density in
the unit cube ${\cal C}_{\delta}=[-1,1]^{\delta}$.
 
Note that ${\cal C}_{\delta}$ is contained within a $\delta$-dimensional
ball of radius $\sqrt \delta$, denoted $\sqrt{\delta} {\cal B}_{\delta}$.
First, we consider points of  ${\cal C}_{\delta}$ as points of
$\sqrt{\delta} {\cal B}_{\delta}$ and apply a homothety with a factor
$\frac{1}{\sqrt{\delta}}$, thereby obtaining
\[
Prob(|p_1,p_2\ldots p_{\delta}|\leq V
                 \mid p_i\in\sqrt{\delta} {\cal B}_{\delta})
=
Prob(|p_1,p_2\ldots p_{\delta}|\leq \sqrt{\delta}^{- \delta} V
                 \mid p_i\in {\cal B}_{\delta})
\leq
         \frac{\sigma_{\delta}}{\sqrt{\delta}^{\delta}} V
\]
 
 Next, we wish to restrict the points to belong to ${\cal C}_{\delta}$, i.e.,
we consider the event $p_i\in\sqrt{\delta} {\cal B}_{\delta}$,
$i=1,\ldots,\delta$, as the union of the
event  $p_i\in{\cal C}_{\delta}$,$i=1,\ldots,\delta$, and its negation.
The probability of this event is clearly
$\left(\frac{2^{\delta}}{v_{\delta}(1) \sqrt{\delta}^{\delta}}\right)^{\delta}$,
so that
 
\[
Prob(|p_1,p_2\ldots p_{\delta}|\leq V
                 \mid p_i\in\sqrt{\delta} {\cal B}_{\delta})
\geq
\left(\frac{2^{\delta}}{v_{\delta}(1) \sqrt{\delta}^{\delta}}\right)^{\delta}
   Prob(|p_1,p_2\ldots p_{\delta}|\leq V
                 \mid p_i\in {\cal C}_{\delta})
\]
\noindent which gives us an upper bound to the probablity in question.
Specifically
 
\[
Prob(|p_1,p_2\ldots p_{\delta} \!|\! \leq V
                 \mid p_i\in {\cal C}_{\delta})
\leq
\left(\frac{v_{\delta}(1) \sqrt{\delta}^{\delta}}{2^{\delta}}\right)^{\delta}
\frac{\sigma_{\delta}}{\sqrt{\delta}^{\delta}} V
\leq
         \frac{\delta v_{\delta}(1) v_{\delta-1}(1)^{\delta}
                        \sqrt{\delta}^{\delta(\delta-1)}}{2^{\delta^2}}
\stackrel{\scriptscriptstyle \Delta}{=}
\psi_{\delta} V
\]

Practically, for small values of $\delta$ we get:

\begin{eqnarray*}
\psi_1       & =    &  1 \\
\psi_2       & =    &                           \pi      \\
\psi_3       & =    &  \frac{27}{128}           \pi^4    \approx 21\\
\psi_4       & =    &  \frac{32}{81}            \pi^6    \approx 380\\
\psi_5       & =    &  \frac{9765625}{402653184}\pi^{12} \approx 23000\\
\psi_6       & =    &  \frac{19683}{125000}     \pi^{15} \approx 4.5\cdot 10^6
\end{eqnarray*}

The previous computation can probably be generalized to other kinds of
domains provided that  the ratio between the volumes of the
inscribed ball and of the
circumscribing ball is bounded.

\subsection{From continuous cube to discrete cube\label{sec_discrete}}
 
 In this section we shall discuss why the obtained
results for continuous density
of points are still useful
for discrete probability, i.e., when points belong to a regular
grid of $\frac{1}{\eta^{\delta}}$ points inside ${\cal C}_{\delta}$.
  
Notice that the preceding results are clearly incorrect for discrete
probabilities. In fact, they prescribe:\\ 
$Prob(|p_1,p_2\ldots p_{\delta}|=0) =0$, which is false for discrete
probabilities.
For example, in two dimensions,
when $p_1$ is chosen, $p_2$ coincides with $p_1$ or with the origin
with probability $2\eta^2$, and yet the determinant value is $0$.
For a less trivial case (when $p_1,p_2$, and the origin are distinct),
the event $p_1=(x_1,x_2)$ and $p_2=(2x_1,2x_2)$, with $x_1<\frac{1}{2}$ and
$y_1<\frac{1}{2}$, has probability $\frac{\eta^2}{4}>0$ while the
determinant is still $0$.
However we can still prove that
$Prob(|p_1\ldots p_{\delta}| \leq V)=O(V)$ when
$\eta$ is smaller than $V$.
 
If $p_1\ldots p_{\delta}$ is a set of points in ${\cal C}_{\delta}$
whose determinant is not larger than $V$, we will map it to
a the nearest set of grid points $p'_1\ldots p'_{\delta}$ whose
determinant is not too large.
More precisely, 
if ${\cal P}=(p_1,\ldots ,p_{\delta})$,
${\cal P'}=(p'_1,\ldots ,p'_{\delta})$
and ${\cal D=P'-P}=(d_1,\ldots ,d_{\delta})$, we have
\[
{\cal |P'|=|P+D|}=|P|+\sum_{I\subset {Z\!\!\!Z}_{\delta},I\neq\emptyset}
        |{\cal PD}_I|
\]
where $I$ is a nonempty subset of $\{1,2\ldots,\delta\}$ and
$|{\cal PD}_I|$ is the determinant obtained by replacing,
for each $i\in I$, $p_i$ with $d_i$ in $|\cal P|$.
The above result follows from the multilinearity of the determinant.
If the cardinality of $I$ is $j$, we can bound from above the absolute value
of $|{\cal PD}_I|$ by the product of the norms of its vector.
Since $||p_i||\leq \sqrt{\delta}$ and $||d_i||\leq \sqrt{\delta}\frac{\eta}{2}$
we have $|{\cal PD}_I|\leq \sqrt{\delta}^\delta \frac{\eta^j}{2^j}$.
By grouping the $\delta \choose j$ 
terms with identical value of $j$ we get:
\[
\left|\;|{\cal P}'| -|{\cal P}|\; \rule[-.3cm]{0cm}{.6cm} \right|
\leq \sqrt{\delta}^\delta \left[ \left(1+\frac{\eta}{2}\right)^{\delta}-1\right]
\approx
\delta \sqrt{\delta}^\delta  \frac{\eta}{2}
\].

 Referring now to the $\delta^{\delta}$-dimensional space whose points
are the sets $p_1\ldots p_{\delta}$, the
determinant $|{\cal P}'|$ is no larger than $V$ if all
the points $\cal P$ in the hypervoxel (in $\delta^{\delta}$ dimensions)
have determinant $|{\cal P}|$ no larger than
$V+\delta \sqrt{\delta}^\delta  \frac{\eta}{2}$.
Since, clearly, a random point $\cal P$ for the continuous distribution
can be in any voxel with the same probability, we conclude that
\begin{small}
\[
Prob\left(\rule[-.3cm]{0cm}{.6cm} |p_1\ldots p_{\delta}| \leq V 
		  \left| \mbox{\scriptsize $
          \begin{array}{c}{\mbox{\scriptsize discrete}}\\
			{\mbox{\scriptsize{distribution in}}}
		  \end{array} $} \right.
		  {\cal C}_{\delta}    \right)
\; \leq \;
Prob\left(\rule[-.3cm]{0cm}{.6cm}
   |p_1\ldots p_{\delta}| \leq V+\delta \sqrt{\delta}^\delta  \frac{\eta}{2}
    	  \left| \mbox{\scriptsize $
          \begin{array}{c}{\mbox{\scriptsize continuous}}\\
			{\mbox{\scriptsize{distribution in}}}
		  \end{array} $} \right.
	{\cal C}_{\delta}    \right)
\] 
\end{small}
\[
\leq\psi_{\delta}.(V+\delta \sqrt{\delta}^\delta\frac{\eta}{2})
\]

\section{The insphere test\label{sec_sphere}}

In the preceding analysis of the "which-side'' predicate,
the points defining the hyperparallelogram were 
assumed to be independent and equally distributed. 
In this section we consider a  
case for which there exist dependencies among the coordinates of the points:
the "insphere'' predicate. This predicate, referred to as $\delta$-insphere for short,
tests whether in $\delta$ dimensions the origin
lies inside the
hypersphere defined by $(\delta+1)$ other arbitrary points
$p_i=(x_{i,1},x_{i,2},\ldots,x_{i,\delta})$, $i=1,2,\ldots,\delta+1$.

It is well known that the $\delta$-dimensional insphere test is embodied
in the sign of the determinant
\[
\begin{array}{cl}
\Delta_{\delta} 
 \stackrel{\scriptscriptstyle \Delta}{=}
   \left| \begin{array}{ccccc} x_{11}&x_{12}& \ldots &x_{11}^2+x_{12}^2+
\ldots+x_{1\delta}^2&1\\
 x_{21}&x_{22}& \ldots &x_{21}^2+x_{22}^2+\ldots+x_{2\delta}^2&1\\
\ldots\\
x_{\delta+2,1}&x_{\delta+2,2}& \ldots &x_{\delta+2,1}^2+x_{\delta+2,2}^2+\ldots+x_{\delta+2,\delta}^2&1\end{array}\right|
\end{array}
\]

Without loss of generality, one of these points ($p_{\delta+2}$) can be chosen
as the origin $O$, so that the above determinant becomes
\[
\Delta_{\delta}
=
   \left| \begin{array}{cccc} x_{11}&x_{12}& \ldots &x_{11}^2+x_{12}^2+
\ldots+x_{1\delta}^2\\
 x_{21}&x_{22}& \ldots &x_{21}^2+x_{22}^2+\ldots+x_{2\delta}^2\\
\ldots\\
x_{\delta+1,1}&x_{\delta+1,2}& \ldots &x_{\delta+1,1}^2+x_{\delta+1,2}^2+\ldots+
x_{\delta+1,\delta}^2\end{array}\right|
\]

These $\delta+1$ points 
are assumed to be evenly distributed in the unit cube ${\cal C}_\delta$.

\subsection{1-insphere test \label{sec_sp1D}}

 We can model the problem as follows. The origin $O$ and $u$ define
 the 1-dimensional sphere; $v$ is the query point. Parameters $u$ and $v$ are independent
 and uniformly distributed in $[-1,1]$, and we wish to evaluate the probability
 of the following event:
  
  \[
  |\Delta_1| \leq A \mbox{\hspace*{3cm}with } \Delta_1=
		\left| \begin{array}{cc} u&u^2\\v&v^2\end{array}\right|
  \]
   
\begin{figure}[t]
\begin{center}
\def\IPEfile{Incircle1.ipe}\input{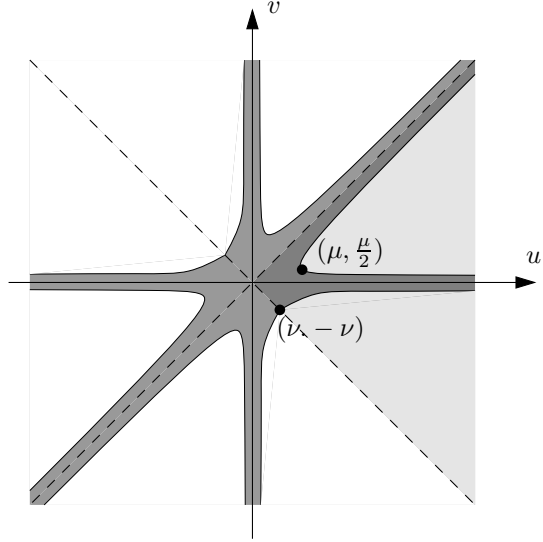}
\caption{The $(u,v)$-region defining $(|\Delta|\leq A)$ in the 
1-insphere test\label{Incircle1}}
\end{center}
\end{figure}

   We can view the above determinant as defined by two points
   $(u,v)$ and $(u^2,v^2)$ in the $u,v$ plane.
   Clearly the choice of point $(u,v)$ completely determines the determinant
   value.
   Point $(u,v)$ is uniformly distributed in the square $[-1,1]\times[-1,1]$.
   Since $uv(v-u)$ is the determinant value, the determinant is null on the
   three lines $u=0$, $v=0$ and $u=v$; its values are symmetric with respect
   to the line $u=-v$ and antisymmetric with respect to $u=v$.
   Therefore for any value of $0\leq A\leq 2$ it is sufficient to evaluate
   the probability in the quadrant $-u\leq v \leq u, 0\leq u\leq 1$
   (fully shaded quadrant in Figure \ref{Incircle1}) and multiply it by 4.
   In the upper semiquadrant (where the determinant is negative)
   the contour lines have equations:
   \[
   v=\frac{u}{2}\pm \frac{1}{2}\sqrt{u^2-4\frac{A}{u}}
   \]
   The two curves for fixed $A$ join with a common vertical tangent
   at the point $(\mu,\frac{\mu}{2})$ where
   $\mu \stackrel{\scriptscriptstyle \Delta}{=} (4A)^{\frac{1}{3}}$
   (notice that such curves exists only for $A<\frac{1}{4})$.
   In the lower semiquadrant (where the determinant is positive)
   contour lines have equations:
   \[
   { v=\frac{u}{2} - \frac{1}{2}\sqrt{u^2+4\frac{A}{u}} }
   \;\;\;\mbox{ and }\;\;\;
   v=-u
   \]
   and intersect the line $v=-u$ at point $ ( \nu , - \nu ) $, where
   $\nu \stackrel{\scriptscriptstyle \Delta}{=} \frac{1}{2}(4A)^{\frac{1}{3}}
   = \frac{1}{2} \mu$.
   The probability of the event described above is given by the heavily shaded
   area in the Figure \ref{Incircle1}
   (in fact this area should be multiplied by 4 since there are four quadrants,
   and normalized, dividing by 4, since 4 is the area of the square).
   The area is also 1 minus the area of the lightly shaded region.
   The latter area, which we want to bound from below is given by:
   \[
   {\cal I} \stackrel{\scriptscriptstyle \Delta}{=}
   \int_{\nu}^1 \left(\frac{3}{2}u -
   \frac{u}{2}\sqrt{1+4\frac{A}{u^3}}\right)du
   +2\int_{\mu}^1\frac{u}{2}\sqrt{1-4\frac{A}{u^3}} du
   \]
   We now observe that
   \[
   \frac{3}{2}-\frac{1}{2}\sqrt{1+4\frac{A}{u^3}} \geq 1 -\frac{A}{u^3}
   \]
   and  (for $A\leq \frac{1}{4}$)
   \[
   \sqrt{1-4\frac{A}{u^3}}\geq 1-4\frac{A}{u^3}
   \]
   so that
    \[
       {\cal I} \geq \int_{\nu}^1 udu
 - A \int_{\nu}^1 \frac{1}{u^2}du
  +  \int_{\mu}^1 udu
-4A \int_{\mu}^1\frac{1}{u^2}du
\] \[
=
 \frac{1}{2}
-\frac{\nu^2}{2}
+A
 -\frac{A}{\nu}
 +\frac{1}{2}
- \frac{\mu^2}{2}
+4A
 -\frac{4A}{\mu}
\]
If we now use the fact the fact that $2\nu=\mu=(4A)^{\frac{1}{3}}$,
then we obtain:
\[
 Prob(|\Delta_1| \leq A)
  \leq
  \frac{17 \sqrt[3]{2}}{4} A^{\frac{2}{3}} -5 A
  \leq
  5.355 A^{\frac{2}{3}}
\]

A direct numerical calculation gives $Prob(|\Delta_1| \leq \frac{1}{4}) =0.7$,
while the value of the above bounding expression for the same value
of $A$ is $0.85$ ( an excellent agreement considering the rather high
value of $A$). For $A\geq \frac{1}{4}$, 
the lightly-shaded region in the upper semi-quadrant disappears yielding
\[
Prob(|\Delta_1| \leq A)
\leq
3\sqrt[3]{2}A^{\frac{2}{3}}
- 4A
\leq
3.78 A^{\frac{2}{3}}
\]
but, considering the high range of $A$, this  expression has little practical
interest.

Alternatively, we may consider the following formulation. Rather than lifting
point $v$ to the parabola $y=v^2u-u^2v$
(for fixed positive $u$), we lift it to the
parabola $y=v^2-uv\stackrel{\scriptscriptstyle \Delta}{=}
F(v)$ (which intersects the $v$-axis at $0$ and $u$) 
so that
\[
Prob(|\Delta| \leq A)= Prob(|F(v)|\leq \frac{A}{u})
\]
This formulation will be useful when considering the multidimensional case in
Section \ref{sec_spdD}

\subsection{Higher dimensional insphere test}

By elementary column operations the determinant $\Delta_{\delta}$
can be transformed
into one where the last column has zero entries except in the last row.
Specifically, $\Delta_{\delta}=0$ defines the circumsphere $S$
of the points $O,p_1,\ldots,p_{\delta}$, whose center is the point
$(\frac{c_1}{2},\ldots,\frac{c_{\delta}}{2})$. Subtracting 
column $i$ times $c_i$ ($i=1,\ldots,\delta$) from the last column, 
we obtain:
\[
\Delta_{\delta} =
   \left| \begin{array}{ccccc} x_{11}&x_{12}& \ldots &x_{1\delta}&0\\
 x_{21}&x_{22}& \ldots & x_{2\delta}&0\\
\vdots& \vdots&        & \vdots     &\vdots\\
x_{\delta 1}&x_{\delta 2}& \ldots &x_{\delta \delta}&0\\
x_{\delta+1,1}&x_{\delta+1,2}& \ldots &x_{\delta+1,\delta}&W\end{array}\right|.
\]
\noindent The determinant of the intersection of the first $\delta$
rows and columns of the above matrix
gives the signed volume $v$ of the hyperparallelogram defined
by the first $\delta$ points and $W$ has 
the dimension of the square of a length and
is the value of the function
\begin{equation}
\label{A5}
F(x_1,\ldots,x_{\delta})=x_1^2+x_2^2+\ldots+x_{\delta}^2 -c_1x_1-\ldots c_{\delta}x_{\delta} 
\end{equation} 
evaluated at point $p_{\delta+1}$. 
We let $ u^2
\stackrel{\scriptscriptstyle \Delta}{=}
\frac{1}{4}(c_1^2+\ldots+c_{\delta}^2)$
denote the square of the radius of sphere $S$. 
Notice also that $x_{\delta+1}=F(x_1,\ldots,x_{\delta})$ is the equation of
a hyperparaboloid ${\cal H }$ in $(\delta+1)$-dimensional space, so that
$F(x_{\delta+1, 1},\ldots,x_{\delta+1,\delta})=F(p_{\delta+1})$ is the signed 
height of the point  obtained by lifting
$p_{\delta+1}$ from the hyperplane $x_{\delta+1}=0$ (to which it
belongs) to ${\cal H}$. Notice that hyperplane $x_{\delta+1}=0$ 
contains the hypersphere $S$. 

We now wish to bound from above the
probability of the event $  |F(p_{\delta+1})| \leq V$, 
for some constant $V$.

Assuming as usual constant density, this probability is the volume
of a sphere $S''$  of radius $u''=\sqrt{V+u^2}$ minus the volume of a
concentric sphere $S'$ of radius $u'=\sqrt{-V+u^2}$,
whenever the latter is defined. (These spheres are intersected with
${\cal C}_{\delta}$ and normalized by its volume.)
In general dimension, since the above radii depend upon the points
$p_1,p_2,\ldots,p_{\delta}$,
the evaluations of the volumes of the above spheres is problematic.
However, we shall show that an interesting simplification occurs for $\delta=2$.
  
Below we shall use the following bounding technique. Let
$\Delta_{\delta}$ be expressed as the product of two continuous random
variables
\[
\Delta_{\delta}= ab
\]
Given a constant $\alpha$ we have:
\begin{eqnarray*}
Prob(ab\leq V)
& \leq  &
Prob(a \leq \alpha \mbox{ or } b \leq \frac{V}{\alpha}) \\
& \leq  &
Prob(a \leq \alpha) + Prob(b \leq \frac{V}{\alpha})  -
Prob(a \leq \alpha \mbox{ and } b \leq \frac{V}{\alpha})
\end{eqnarray*}
\noindent that is:
\begin{eqnarray}
\label{A6}
Prob(ab\leq V) &\leq& Prob(a \leq \alpha) + Prob(b \leq \frac{V}{\alpha})
\end{eqnarray}

\subsubsection{2-insphere test}
  
  In this case the volumes of $S''$ and $S'$ are respectively
  $\pi(V+u^2)$ and $\pi(-V+u^2)$,
  so that, if $S'$ is defined, then their difference becomes
  $\pi 2V$, otherwise (i.e., when $u^2 \leq V$) $\pi(V+u^2)\leq \pi 2V$.
  After normalization (since 4 is the measure of the unit square) we have
  \[
  Prob (|F(p_3)|\leq V)\leq \frac{\pi V}{2}.
  \]
  \noindent Given that $|\Delta_2|=|p_1p_2| F(p_3)$
  inequality (\ref{A6}) becomes
  \begin{eqnarray*}
  Prob(| \Delta_{2} |\leq V) 
  &\leq&
   Prob(| p_1 p_2| \leq \alpha)+ Prob( |F(p_3)| \leq \frac {V}{\alpha})\\
   &\leq &
   \psi_2 \alpha + \frac{\pi V}{2\alpha}.
   \end{eqnarray*}

   \noindent
   From Section 3 we know that $\psi_2=\pi$. Selecting for
$\alpha$ the critical value  
   $\alpha=\sqrt{\frac{V}{2}}$ we have
	\[
	Prob(| \Delta_{2} |\leq V)
               =\pi \sqrt{2V} \approx  4.44\sqrt{V}
	\]

\subsubsection{$\delta$-insphere test ($\delta >2$) \label{sec_spdD}}

As for the two-dimensional case, we shall prove that, in general, the
probability that $|F(p_{\delta+1})|$ is no larger than $W$ is
sufficiently small.

Again, inequality (\ref{A6}) becomes:
\begin{equation}
\label{A7}
 Prob(| \Delta_{\delta} |\leq W)
 \leq
 Prob(| p_1\ldots p_{\delta}| \leq \frac{W}{\alpha})
 + Prob( |F(p_{\delta +1})| \leq \alpha)
\end{equation} 
 \noindent
 We know, from the results of Section \ref{sec_cube}, that
 \[
 Prob(| p_1\ldots p_{\delta}| \leq \frac{W}{\alpha})
 \leq \psi_{\delta}\frac{W}{\alpha}
 \]

\begin{figure}[t]
\begin{center}
\def\IPEfile{Pstar.ipe}\input{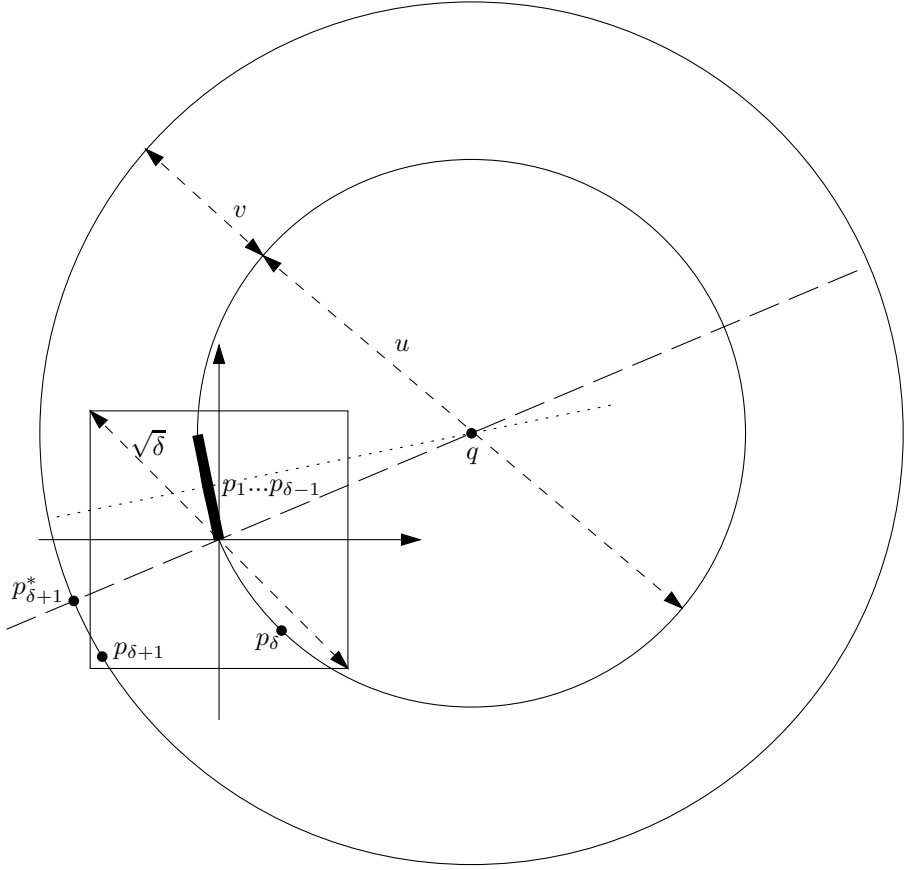}
\caption{Definition of $p_{\delta +1}^*$, $u$ and $v$, using a 
2-dimensional instantiation\label{Pstar}}
\end{center}
\end{figure}

Therefore, there remains to bound from above
$Prob( |F(p_{\delta +1})| \! \leq \!  \alpha)$.
Recall that $ |F(p_{\delta +1})|$
is the distance from the plane $x_{\delta+1}=0$ of the point $p_{\delta+1}$
lifted to the hyperparaboloid ${\cal H}$, of equation $F(p)=0$, which intersects
$x_{\delta+1}=0$ in a $\delta$-dimensional sphere passing by the origin with
with center $q=(\frac{c_1}{2},\ldots,\frac{c_{\delta}}{2})$.
We define  point $p_{\delta +1}^*$ such that
it lies on a line $l$ passing by $O$ and $q$ and
such that $\mbox{length}(p_{\delta +1}^*,q)=\mbox{length}(p_{\delta +1},q)$
and among the two possible choices, we select the one closest to  $O$
(see Figure \ref{Pstar}).
It is immediate that $ |F(p_{\delta +1})|= |F(p_{\delta +1}^*)|$ and 
that $\mbox{length}(Op_{\delta +1}^*) \leq \sqrt{\delta}$.

Thus, our problem is reduced to a one-dimensional instance,
closely related to the one we studied in Section \ref{sec_sp1D}
(here  $u=2 \mbox{length}(Oq)$ and $v$ is the signed value of
$\mbox{length}(Op_{\delta +1}^*)$).
There are
however, some significant differences. There, we were evaluating the probability
of the event $|uF(v)| \leq V$ , and variables $u$ and $v$ were uniformly
distributed in $[-1,1]$.
Here, on the other hand, we wish to evaluate the probability
of the event $|F(v)| \leq \alpha$, ``radius'' $u$
varies between $0$ and $\infty$,
and 
``distance'' $v$ varies between $-\sqrt{\delta}$ and $\sqrt{\delta}$.
Moreover, the densities $p_1(u)$ and $p_2(v)$ of $u$ and $v$ respectively
are not constant; however, as we shall see,
they are bounded by constants {$q_1$ and $q_2$.

Next, we observe that
\begin{eqnarray*}
Prob(|F(p_{\delta+1})| \leq \alpha)
&=&
\int_{u=0}^{\infty} \int_{v=-\sqrt{\delta}}^{\sqrt{\delta}} p_1(u) p_2(v)
		Prob(|F(p_{\delta+1})| \leq \alpha \mid u,v) dv du\\
& =&
2\int_{u=0}^{\infty} \int_{v=0}^{\sqrt{\delta}} p_1(u) p_2(v)
				Prob(|F(p_{\delta+1})| \leq \alpha \mid u,v) dv du
\end{eqnarray*}
\noindent
Since
the function $F(p_{\delta +1})$ has the expression $v^2-uv$,
the right-hand-side of the above equation
is just the integral of $p_1(u) p_2(v)$
on the domain( shown in Figure \ref{Hyperbole}) bounded 
by the curves $v^2-uv=\pm\alpha$.
\begin{figure}[tp]
\begin{center}
\def\IPEfile{Hyperbole.ipe}\input{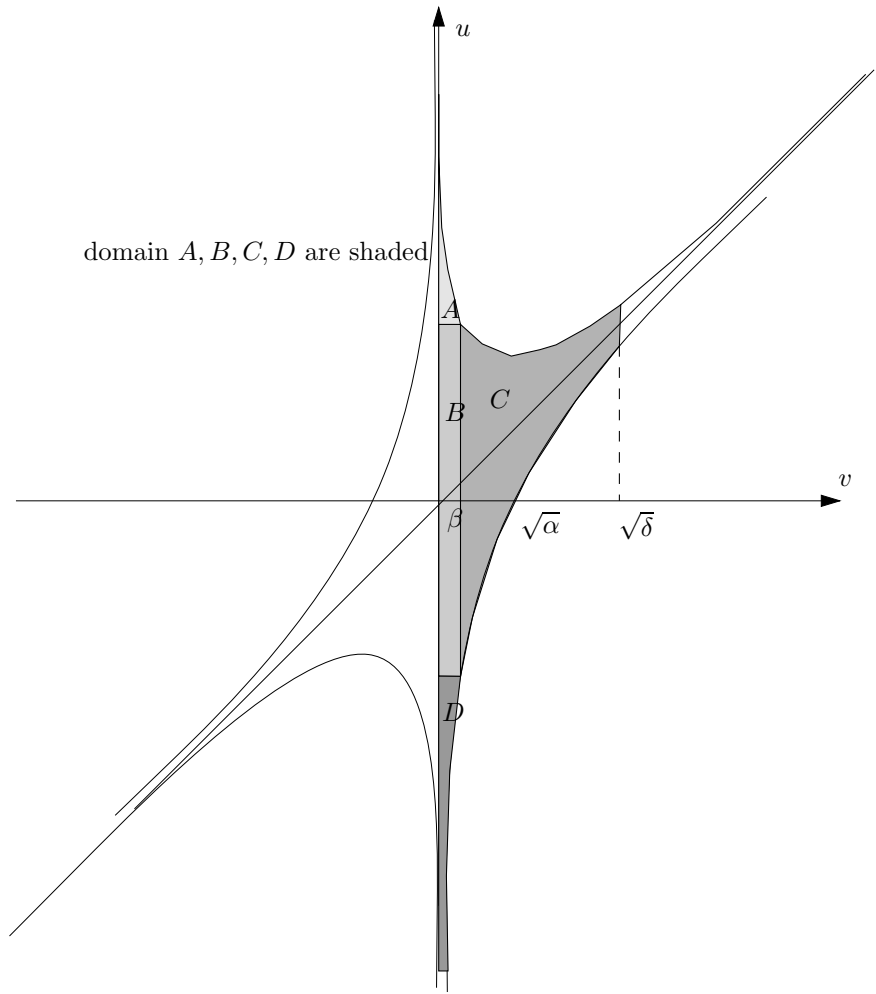}
\caption{Integration domain\label{Hyperbole}}
\end{center}
\end{figure}
This integration domain can be split into four subdomains
A,B,C,D, as shown in Figure \ref{Hyperbole}.
This split  depends on a parameter $\beta \leq \sqrt{\alpha}$
which will be chosen later. In detail we have:

\begin{itemize}
\item Domain A is defined by
$ \beta\leq v\leq \sqrt(\delta)$ and
$v-\frac{\alpha}{v} \leq u \leq v+\frac{\alpha}{v}$.
$p_1(u)p_2(v)\leq q_1q_2$ and thus
\[
\int\!\!\!\int_A p_1(u)p_2(v) dv du
\leq
q_1 q_2 \int_\beta^{\sqrt{\delta}} \frac{2\alpha}{v} dv
\leq
q_1 q_2 \left(\ln \delta + 2 \ln \frac{1}{\beta} \right)\alpha
\]
\item Domain B is a rectangle defined by
$ 0\leq v \leq \beta$ and
$\beta-\frac{\alpha}{\beta} \leq u \leq \beta+\frac{\alpha}{\beta}$.
$p_1(u)p_2(v)\leq q_1q_2$ and thus
\[
\int\!\!\!\int_B p_1(u)p_2(v) dv du
\leq
q_1 q_2 \beta \frac{2\alpha}{\beta}
\leq
2 q_1 q_2 \alpha
\]
\item Domain C is defined by
$ 0\leq v \leq \beta$ and
$\beta+\frac{\alpha}{\beta} \leq u \leq v+\frac{\alpha}{v}$.
$p_2(v)\leq q_2$ and since $\int p1(u)du\leq 1$ on any domain,
we get
\[
\int\!\!\!\int_C p_1(u)p_2(v) dv du
\leq
q_2 \beta
\]
\item Domain D is analogous to domain C,
\[
\int\!\!\!\int_D p_1(u)p_2(v) dv du
\leq
q_2 \beta
\]
\end{itemize}

These results yield:
\begin{equation}
\label{A8}
Prob(|F(p_{\delta+1})| \leq \alpha)
=
2q_1 q_2 \alpha\left(\ln \delta + 2 \ln \frac{1}{\beta} + 2\right) +4q_2 \beta.
\end{equation}

To obtain values for $q_1$ and $q_2$, we observe:
\begin{enumerate}
\item  With reference to  $p_1(u)$, consider
the $(\delta\!-\!1)$-dimensional sphere passing
by $O,p_1\!\ldots p_{\delta \!-\!1}$ in the flat $\Pi$ defined by these points
(see Figure \ref{P1} for an illustration), 
and let $t$ be its center.
Consider the family ${\cal F}$ of $\delta$-dimensional spheres
passing through the origin and whose
centers project to $t$ in $\Pi$. Point $p_{\delta}$ will determine a
$\delta$-dimensional sphere of radius between $u$ and $u+du$ if it lies
between a pair of spheres of $\cal F$ (or its symmetric pair with
respect to $\Pi$) of radii  $u$ and $u+du$. The volume $p_1(u)du$ of this
region (normalized by the volume $2^{\delta}$ of
${\cal C}_{\delta}$) is $\leq \frac{2}{2^{\delta}}{\cal A}_{\delta} du$,
where ${\cal A}_{\delta}$ is the maximum of the measure of the surface
of a $\delta$-dimensional sphere passing by the origin and intersected with
${\cal C}_{\delta}$. A trivial upper bound on ${\cal A}_{\delta}$
is the measure $2\delta 2^{\delta-1}$ of the surface of ${\cal C}_{\delta}$.
\begin{figure}[t]
\begin{center}
\def\IPEfile{P1.ipe}\input{P1.ipe}
\caption{For the analysis of $p_1(u)$\label{P1}}
\end{center}
\end{figure}
Therefore
\[
p_1(u) \leq \frac{2}{2^{\delta}}.2\delta 2^{\delta-1}= 2\delta
\stackrel{\scriptscriptstyle \Delta}{=} q_1
\]
\item With reference to $p_2(v)$, we observe that once
$p_{\delta}$ is fixed, the choice of $p_{\delta+1}$ will
produce  a value between $v$ and $v+dv$ if $p_{\delta+1}$
belongs to the shaded region in Figure \ref{P2}.
The volume of this region is $dv$ times the surface of the
sphere of radius $v$ inside ${\cal C}_{\delta}$. This surface
is clearly less than the surface of ${\cal C}_{\delta}$, so that we get
\begin{figure}[t]
\begin{center}
\def\IPEfile{P2.ipe}\input{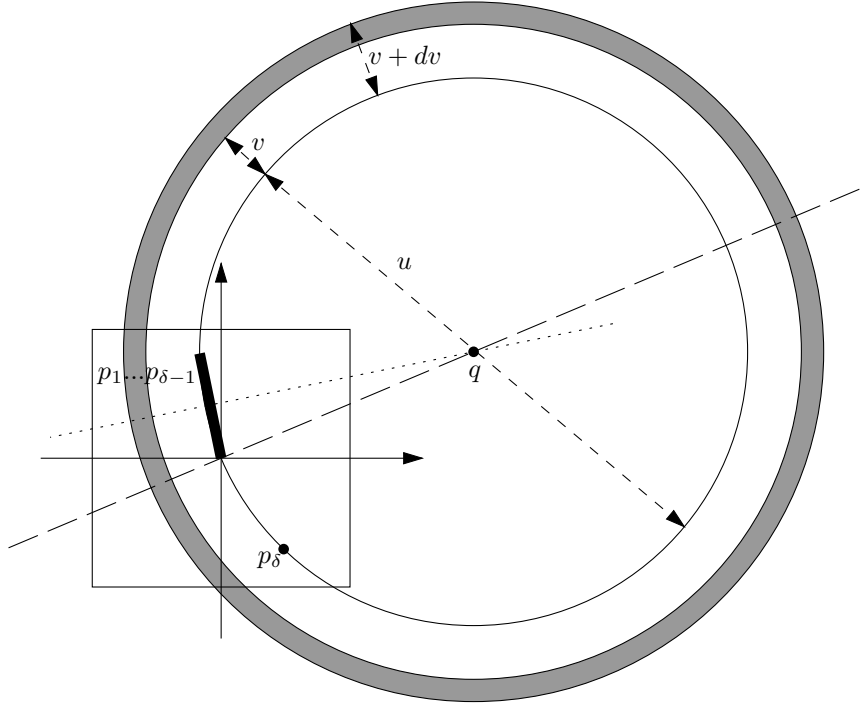}
\caption{For the analysis of $p_2(v)$\label{P2}}
\end{center}
\end{figure}
\[
p_2(v) \leq \frac{1}{2^{\delta}}.2\delta 2^{\delta-1}= \delta
\stackrel{\scriptscriptstyle \Delta}{=} q_2
\]
\end{enumerate}

Substituting $q_1=2\delta$ and $q_2=\delta$ into (\ref{A8})
and choosing $\beta=\frac{\alpha}{2}<\sqrt{\alpha}$ 
we obtain:
\[
Prob(|F(p_{\delta+1})| \leq \alpha)
\leq
8 \delta^2 \alpha \ln \frac{1}{\alpha}
+
\left(4 \delta^2 \ln \delta + 8 \delta^2 \ln 2
         + 8 \delta^2
		 + \delta
\right)\alpha
\stackrel{\scriptscriptstyle \Delta}{=}
\tau_{\delta} \alpha \ln\frac{1}{\alpha} + \theta_{\delta} \alpha
\]
(Small values of $\tau_{\delta}$ and $\theta_{\delta}$ are
$\tau_3= 72$,    $\theta_3=164.5$,
$\tau_4= 128$,   $\theta_4=309.4$,
$\tau_5=200$,    $\theta_5=504.6$,
$\tau_6=288$ and $\theta_6=751.6$.)
Inequality (\ref{A7}) becomes:
\[
Prob(| \Delta_{\delta} |\leq W)
\leq
 \psi_{\delta}\frac{W}{\alpha}
 + \tau_{\delta} \alpha \ln\frac{1}{\alpha} + \theta_{\delta} \alpha
\]
so that, setting $\alpha=\sqrt{\frac{\psi_{\delta}W}{\tau_{\delta}+\theta_{\delta}}}$
\[
Prob(| \Delta_{\delta} |\leq W)
\leq
 \sqrt{\psi_{\delta} (\tau_{\delta}+\theta_{\delta})} \sqrt{W}
	 \left(    \ln\frac{1}{W}
			 - \ln\frac{\tau_{\delta}+\theta_{\delta}}{\psi_{\delta}}
			 + 1 \right)
\]
\[
\hspace*{3cm}
\stackrel{\scriptscriptstyle \Delta}{=}
\phi_{\delta} \sqrt{W}\ln\frac{1}{W} + \chi_{\delta} \sqrt{W}
\]

\begin{eqnarray*}
\phi_3 \approx 70    & & \chi_3 = -100\\
\phi_4 \approx 408   & & \chi_4 =  350\\
\phi_5 \approx 3970  & & \chi_5 = 18000\\
\phi_6 \approx 68500 & & \chi_6 = 640000
\end{eqnarray*}

\subsection{Discrete distribution}

As for the case of the which-side test, also for the insphere test
we can map each point 
to the nearest grid point, and evaluate the ensueing  effect on the
determinant to be computed.
Generalizing the notation of Section \ref{sec_discrete},
we have that  the norm  $||p^{\dag}_i||$ of the lifted point is bounded
by $\sqrt{\delta+\delta^2}$
and that 
$||d^{\dag}_i|| \leq \sqrt{\delta} \sqrt{2\frac{\eta^2}{4} + 2\frac{\eta}{2}}
\leq \frac{\sqrt{\delta}}{\sqrt{2}} \sqrt{\eta+\eta^2}$.
Therefore, grouping the errors term, we get
\[
\left|\;|{\cal P}'^{\dag}| -|{\cal P^{\dag}}|\; \rule[-.3cm]{0cm}{.6cm} \right|
\leq 
\left[ \sqrt{\delta+\delta^2}
		+ \frac{\sqrt{\delta}}{\sqrt{2}} \sqrt{\eta+\eta^2} \right]^{\delta+1}
- \sqrt{\delta+\delta^2}^{\delta+1}
\] \[ \hspace*{5cm}
\raisebox{-.3ex}{$\stackrel{<}{\sim}$}
(\delta+1) \sqrt{\delta+\delta^2}^{\delta+1} \cdot
  \frac{\sqrt{\delta}}{\sqrt{2}\sqrt{\delta+\delta^2}} \sqrt{\eta+\eta^2}
\]
and considering $\eta$ small
\[
\left|\;|{\cal P}'^{\dag}| -|{\cal P^{\dag}}|\; \rule[-.3cm]{0cm}{.6cm} \right|
\raisebox{-.3ex}{$\stackrel{<}{\sim}$}
(\delta +1 ) \sqrt{\frac{\delta}{2}}\sqrt{\delta+\delta^2}^{\delta}
  \cdot\sqrt{\eta}
\]

\section{The efficacy of arithmetic filters}
To complete the analysis, in this section we wish to assess, under the
given statistical assumptions, the probability of failure of a given
arithmetic filter for determinant sign evaluation (i.e., the probability 
that the filter is unable to certify the correctness of the computed sign).
This probability is a quantitative measure of the efficacy of
the filter.  If $|p_1\ldots p_{\delta}|>
\varepsilon_{\delta}$, then the result of the computation is reliable, and so
is its sign. Plugging $\varepsilon_{\delta}$ in place of $V$ in the
expression above, we obtain the condition:
\[
Prob\left(\rule[-.3cm]{0cm}{.6cm} |p_1\ldots p_{\delta}| \leq
												  \varepsilon_{\delta}
                  \left| \mbox{\scriptsize $
          \begin{array}{c}{\mbox{\scriptsize discrete}}\\
                        {\mbox{\scriptsize{distribution in}}}
                  \end{array} $} \right.
                  {\cal C}_{\delta}    \right)
\;\; \leq \;\;
\psi_{\delta}.(\varepsilon_{\delta}+\delta \sqrt{\delta}^\delta
 \frac{\eta}{2})
 \stackrel{\scriptscriptstyle \Delta}{=}\rho_{\delta}
 \]
 where we have used the result (Section 3) that
 $Prob(|p_1\ldots p_{\delta}| \leq V)\leq \psi_{\delta}V$. The parameter
$\rho_{\delta}$ introduced here is therefore the sought measure
of filter efficacy.

To exemplify this approach, we shall compute $\rho_{\delta}$ for the 
evaluation of determinants, for the case where the coordinates
 of the points are floating-point numbers in the interval  $[-1,1]$,
the computations are carried out using floating-point
arithmetic with $b$ bits of mantissa, and the determinant is
evaluated by standard expansion with respect to one of its columns
(recursive evaluation).

To this end, it is necessary to compute the parameter $\varepsilon_{\delta}$.
We introduce the following notation:
 
\begin{itemize}
\item  $\P{M}{m}$  denotes the set of numbers
whose absolute value is bounded by $M$
and whose error is bounded by $m$.
Original entries belong to $\P{1}{0}$.
\item $\overline{M}$ denotes $2^{\lceil \log M \rceil}$.
\end{itemize}
With this notation, if $x_1\in\P{M_1}{m_1}$ and $x_2\in\P{M_2}{m_2}$, then we
have:
\[
x_1+x_2\in\P{M_1+M_2}{2^{-b-1}\cdot\overline{ M_1+M_2}
+ m_1 + m_2}
\]
\[
x_1\cdot x_2\in\P{M_1\cdot M_2}
{2^{-b-1}\cdot\overline{ M_1\cdot M_2} +m_1\cdot M_2 + m_2\cdot M_1}
\]
\noindent  These rules express the mechanics of $b$-bit mantissa
normalizing floating-point operations with round-off
(round-off is done to the nearest, thus the error done is half of the value
of the last bit).
 After transforming
variable  $y$ to an $\P{M}{m}$ pair, we shall
express the above rules as an arithmetics on such pairs, as follows:
\begin{enumerate}
\item $\P{M_1}{m_1} + \P{M_2}{m_2}= \P{M_1 +M_2}{2^{-b-1}\cdot\overline{ M_1+M_2
}
+ m_1 + m_2}$
\item $\P{M_1}{m_1} \cdot  \P{M_2}{m_2}=\P{M_1\cdot M_2}
{2^{-b-1}\cdot\overline{ M_1\cdot M_2} +m_1\cdot M_2 + m_2\cdot M_1}$
\end{enumerate}

The results given below are obtained in an Appendix to this paper
by the mechanical application (with a few noted exceptions)
of the above two rules to the recursive  evaluation of a 
determinant. Using a $b$-bit mantissa, the results are:
\begin{itemize}
\item      \makebox[4cm][l]{$\varepsilon_2 = { 2     \cdot 2^{-b}} $ }
 $\bullet$ \makebox[4cm][l]{$\varepsilon_5 = {576    \cdot 2^{-b}} $ }
 $\bullet$ \makebox[4cm][l]{$\varepsilon_8 = {226624 \cdot 2^{-b}} $ }
\item      \makebox[4cm][l]{$\varepsilon_3 = {13     \cdot 2^{-b}} $ }
 $\bullet$ \makebox[4cm][l]{$\varepsilon_6 = {3672   \cdot 2^{-b}} $ }
\item      \makebox[4cm][l]{$\varepsilon_4 = {76     \cdot 2^{-b}} $ }
 $\bullet$ \makebox[4cm][l]{$\varepsilon_7 = {27304  \cdot 2^{-b}} $ }
\end{itemize}

For example, using the IEEE norm on $b=53$ bits, we can therefore estimate the
corresponding probablility of failure. The pertinent values of
 $\varepsilon_{\delta}$, $\delta\sqrt{\delta}^\delta\frac{\eta}{2}$, and
 $\rho_{\delta}$ are displayed below:
\begin{itemize}
\item \makebox[4cm][l]{$\varepsilon_2 = { 2.2\cdot 10^{-16}} $ }
 $\bullet$ \makebox[4cm][l]{$\delta\sqrt{\delta}^\delta\frac{\eta}{2}=
                  1.6 \cdot 10^{-16} $}
 $\bullet$ $\rho_2= { 1.2\cdot 10^{-15}} $
\item \makebox[4cm][l]{$\varepsilon_3 = { 1.4\cdot 10^{-15}} $ }
 $\bullet$ \makebox[4cm][l]{$\delta\sqrt{\delta}^\delta\frac{\eta}{2}=
                  8.7 \cdot 10^{-16} $}
 $\bullet$ $\rho_3= { 4.8\cdot 10^{-14}} $
\item \makebox[4cm][l]{$\varepsilon_4 = { 8.4\cdot 10^{-15}} $ }
 $\bullet$ \makebox[4cm][l]{$\delta\sqrt{\delta}^\delta\frac{\eta}{2}=
                  7.1 \cdot 10^{-15} $}
 $\bullet$ $\rho_4= { 5.9\cdot 10^{-12}} $
\item \makebox[4cm][l]{$\varepsilon_5 = { 5.7\cdot 10^{-14}} $ }
 $\bullet$ \makebox[4cm][l]{$\delta\sqrt{\delta}^\delta\frac{\eta}{2}=
                  7.8 \cdot 10^{-14} $}
 $\bullet$ $\rho_5= { 3.0\cdot 10^{-9}} $
\item \makebox[4cm][l]{$\varepsilon_6 = { 8.3\cdot 10^{-13}} $ }
 $\bullet$ \makebox[4cm][l]{$\delta\sqrt{\delta}^\delta\frac{\eta}{2}=
                  1.1 \cdot 10^{-12} $}
 $\bullet$ $\rho_6= { 8.7\cdot 10^{-6}} $
\end{itemize}

Finally, it is important to evaluate the efficiency of the examined
filter. We observe here that a recursive evaluation uses $\delta !$
operations, but the cost can be reduced using dynamic programming.  In
fact, the recursive evaluation involves $\delta \choose i$ minors of
dimension $i$.  In turn each such minor involves $(2i-1)$ arithmetic
operations ($i$ multiplication and $i-1$ additions), whose operands
are either minors of smaller dimension or original coefficients (for
$i=2$, where the recursion stops). Thus, the total number of
operations is: \[ r_{\delta}=
\sum_{0}^{\delta -2} {\delta \choose i}(2(\delta -j)-1)
=(\delta -1)(2^{\delta}-1) \sim \delta2^{\delta}.  \]
 
For $\delta \leq 7$ we obtain
\noindent $r_1 = 0$, $r_2= 3$,$r_3 = 14$, $r_4 = 45$, $r_5 = 124$, $r_6 = 315$,$
r_7 =762$ and $r_8=1785$.

It must be pointed out that, for the same function considered above
(determinant value), one may choose alternative evaluation schemes
(corresponding to different expressions of the given function, such as
other expansion rules, Gaussian elimination, etc.)  and/or different
arithmetic engines. Each such choice would embody a filter, whose
efficacy and efficiency can be assessed with the outlined method.

\section{ Summary of results and conclusions}
In this paper we have developed a general approach to the assessment
of the efficacy of arithmetic filters, under some reasonable
probability assumptions. As an important example, we have considered
the efficacy of filters for the evaluation of signs of determinants,
both for a case where all entries are independent (the "which-side''
predicate) and for a case where dependencies exist (the "insphere''
predicate). This analysis, in general, consists of two parts.

The first part aims at computing the threshold for certification by the
filter (i.e., the maximum error which can be generated by the
evaluation process), and does not rest on any assumption about the
distribution of the data. As an example, we have carried out this threshold
analysis for the so-called recursive evaluation procedure, which computes
a determinant by expanding it with respect to one of its columns.
 
The second part, which is considerably more subtle, aims at establishing the
probability of failure of the filter, i.e., the probability that the
result of the computation falls below the threshold. This analysis rests
on {\em a priori} assumptions on the distribution of the input data,
which we have taken as uniform within their representation
range, and has been carried out for the two important geometric tests
mentioned above. With the notations introduced in the preceding sections,
the results are summarized below.

\begin{footnotesize}
\begin{eqnarray*}
Prob(|p_1,p_2\ldots p_{\delta}|\leq V \mid p_i
				   \mbox{ continuous in } {\cal B}_{\delta})
& \leq &
\sigma_{\delta} V\\
Prob(|p_1,p_2\ldots p_{\delta}|\leq V \mid p_i
				\mbox{ continuous in } {\cal C}_{\delta})
& \leq &
\psi_{\delta} V\\
Prob(|p_1,p_2\ldots p_{\delta}|\leq V \mid p_i
		\in \mbox{ $\eta$-grid in } {\cal C}_{\delta})
& \leq &
\psi_{\delta}.(V+\alpha_{\delta} \eta)\\
Prob(|\Delta_1|\leq W \mid p_i
				\mbox{ continuous in } {\cal C}_{1})
& \leq &
5.36 W^{\frac{2}{3}}\\
Prob(|\Delta_{\delta}|\leq W \mid p_i
									\mbox{ continuous in } {\cal C}_{\delta})
& \leq &
\phi_{\delta} \sqrt{W}\ln\frac{1}{W} + \chi_{\delta} \sqrt{W} \\
Prob(|\Delta_i|\leq W \mid p_i
								\in \mbox{ $\eta$-grid in } {\cal C}_{\delta})
& \leq &
\phi_{\delta} \sqrt{W + \beta_{\delta} \sqrt{\eta}}{\ln\frac{1}{W + \beta_{\delta} \sqrt{\eta}}} + \chi_{\delta} \sqrt{W + \beta_{\delta} \sqrt{\eta}} \\
\end{eqnarray*}
\end{footnotesize}

Below we repeat for synoptic convenience the definitions of the relevant 
constants and a tabulation of their values for small $\delta$.

\begin{eqnarray*}
\sigma_{\delta}
& = &
\delta \frac{ v_{\delta-1}(1)^{\delta} }{ v_{\delta}(1)^{\delta -1}} \\
\psi_{\delta}
& = &
\frac{\delta v_{\delta}(1) v_{\delta-1}(1)^{\delta}
                        \sqrt{\delta}^{\delta(\delta-1)}}{2^{\delta^2}} \\
\phi_{\delta}
& = &
\sqrt{\psi_{\delta}
 \left(2 \delta^2 \ln \delta + 4 \delta^2 \ln 2 + 8 \delta^2 + \frac{\delta}{2}
 \right)} \\
\chi_{\delta}
& = &
\phi{\delta} \left( 1 - \ln \frac{
 2 \delta^2 \ln \delta + 4 \delta^2 \ln 2 + 8 \delta^2 + \frac{\delta}{2}
 }{\psi_{\delta}}
 \right) \\
\alpha_{\delta}
& = &
\frac{\delta \sqrt{\delta}^\delta}{2}\\
\beta_{\delta}
& = &
(\delta+1) \sqrt{\frac{\delta}{2}} \sqrt{\delta+\delta^2}^{\delta}
\end{eqnarray*}

The values for small $\delta$ are

\vspace*{.25in}

\begin{tabular}{|c|c|c|c|c|c|c|}
\hline
$\delta$&$\sigma_{\delta}$&$\psi_{\delta}$&$\phi_{\delta}$&$\chi_{\delta}$&$\alpha_{\delta}$&$\beta_{\delta}$\\
\hline
1     & 1     & 1               &     &      & 0.5   & 2      \\ \hline
2     & 2.5   & 3.2             &     & 4.4  & 2     & 18      \\ \hline
3     & 5.3   & 21              & 70  &-100  & 7.8   & 200      \\ \hline
4     & 10    & 380             & 408 & 350  & 32    & 2800      \\ \hline
5     & 19    & 23000           & 3970&18000 & 140   & 47000      \\ \hline
6     & 35    & $4.5\cdot 10^6$ &68500&640000& 648   & 900000      \\ \hline
\end{tabular}
\vspace*{.25in}

The above values are tight only for the constant $\sigma$. Due to data 
interdependencies, the analysis of the insphere predicate is considerably more 
involved than that of the which-side predicate, and the adverse effect of 
the dependencies is manifest in the larger values of the
probability of failure.

Such analysis is particularly valuable for estimating the time required 
to test the determinant sign. Under the given probability assumptions, 
we may conclude that for small dimension ($\leq 6$) straightforward
floating-point filters (i.e., floating-point evaluators) are 
extraordinarily effective.

\newcommand{\etalchar}[1]{$^{#1}$}

\section{Appendix. Error evaluation\label{application}}

$D_i$ represents a generic determinant of dimension $i$,
and $x$ a generic original coordinate.
 It must be pointed out that the recursive evaluation technique yields an upper
bound of $\delta !$ for $D_{\delta}$ 
because dependencies among data are not exploited.
However,smaller values of such bound are known:
Since the determinant value is bounded by the product 
of the norms of its $\delta$ components,
$\sqrt{\delta}^{\delta}$ is an upper bound on $D_{\delta}$,
which is attained when $\delta$ is a power of two (Hadamard matrices).
For small values of $\delta$ the following bounds have 
been obtained ( some by exhaustive calculation):  
$D_{1}\leq 1$,
$D_{2}\leq 2$,
$D_{3}\leq 4$,
$D_{4}\leq 16$,
$D_{5}\leq 48$,
$D_{6}\leq 160$,
$D_{7}\leq 576$ and
$D_{8}\leq 4096$.
Whenever applicable, we shall use these results 
below to obtain tighter estimates. These estimates have the form
\[ D_{\delta} \in \P{G_{\delta}}{\varepsilon_{\delta}} 
\]
where $\varepsilon_{\delta}$ is the object of the analysis and 
$G_{\delta}$, the largest value attainable by $D_{\delta}$, is
only needed to carry out the analysis.

\begin{enumerate}
\item $D_{1}\in\P{1}{0}$, by definition.

\item $D_{2}\in\P{2}{2^{-b+1}}$. In fact $D_2= x \cdot D_1 + x \cdot D_1$,
$x \in \P{1}{0}$ and $D_1 \in \P{1}{0}$. Therefore
\begin{eqnarray*}
      \P{1}{0} \cdot \P{1}{0} + \P{1}{0}\cdot\P{1}{0}     &=&
      \P{1}{2^{-b-1}}+\P{1}{2^{-b-1}} = \P{2}{2^{-b+1}}
\end{eqnarray*}

\item $D_{3}\in\P{4}{13 \cdot 2^{-b}}$. In fact,
$D_3= (x\cdot D_2+x \cdot D_2)+x \cdot D_2$. Therefore:
\begin{eqnarray*}
D_3 & \in & ( \P{1}{0} \cdot \P{2}{2^{-b+1}}+ \P{1}{0} \cdot \P{2}{2^{-b+1}})+
 \P{1}{0} \cdot \P{2}{2^{-b+1}}
\\ & &
 =\P{4}{ 2^{-b-1}\cdot 4+6\cdot 2^{-b}}+\P{2}{3 \cdot 2^{-b}}
\\ & &
 =\P{4}{ 2^{-b-1}\cdot 4 + 11\cdot 2^{-b}}= \P{4}{13 \cdot 2^{-b}}
\end{eqnarray*}

where we have used the fact $D_{4}\leq 4$.

\item $D_{4}\in\P{16}{19\cdot 2^{-b+2}}$. The result is obtained by 
applying the previous rules and $x \in \P{1}{0}$, 
$D_{3}\in\P{4}{13 \cdot 2^{-b}}$ to the following evaluation scheme:
\[
D_4= (x\cdot D_3+x \cdot D_3)+(x \cdot D_3 +x \cdot D_3). 
\]

\item $D_{5}\in\P{48}{129\cdot 2^{-b+2}}$. The result is obtained by 
applying the previous rules and $x \in \P{1}{0}$,
$D_{4}\in\P{16}{19\cdot 2^{-b+2}}$ to the following evaluation scheme:
\[
D_5= ((x \cdot D_4+x \cdot D_4)+x \cdot D_4)+(x \cdot D_4+x \cdot D_4).
\] 
making also use of the fact $D_{5}\leq 48$.

\item $D_{6}\in\P{160}{459\cdot 2^{-b+3}}$. The result is obtained by 
applying the previous rules and $x \in \P{1}{0}$,
$D_{5}\in\P{48}{125\cdot 2^{-b+2}}$ to the following evaluation scheme:
\[
D_6= (((x \cdot D_5+x \cdot D_5)+(x \cdot D_5+x \cdot D_5))+(x \cdot D_5+x \cdot D_5))
\]
making also use of the fact $D_{6}\leq 160$.

\item $D_{7}\in\P{576}{3413\cdot 2^{-b+3}}$. The result is obtained by 
applying the previous rules and $x \in \P{1}{0}$,
$D_{6}\in\P{160}{459\cdot 2^{-b+3}}$ to the following evaluation scheme:
\[
D_7= (((x \cdot D_6+x \cdot D_6)+(x \cdot D_6+x \cdot D_6))+((x \cdot D_6+x \cdot D_6)+x \cdot D_6))
\]
making also use of the fact $D_{7}\leq 576$.

\item $D_{8}\in\P{4096}{3541\cdot 2^{-b+6}}$. Again, the result is obtained by 
applying the previous rules and $x \in \P{1}{0}$,
$D_{7}\in\P{576}{3413\cdot 2^{-b+3}}$ to the following evaluation scheme:
\[
D_8= (((x \cdot D_7+x \cdot D_7)+(x \cdot D_7+x \cdot D_7))+((x \cdot D_7+x \cdot D_7)+(x \cdot D_7+x \cdot D_7)))
\]
making also use of the fact $D_{8}\leq 4096$.

\end{enumerate}

\noindent  REMARK. If the original values do not belongs to $\P{1}{0}$
 but to $\P{1}{\epsilon}$ and $\epsilon$ is small enough, then the
 final error on $D_{i}$ must be increased by $i!\epsilon$.  $\epsilon$
 small enough means that the error on $M$ does not affect the value
 $\overline{M}$, in particular if all the manipulated sets are of the
 form $\P{M}{m}$ with $\overline{M}=\overline{M+m}$, then adopting a
 first-order approximation is legitimate.

\end{document}